\documentstyle[amsfonts,amsmath,epsfig,12pt]{article}

\newcommand {\slsh} [1] {\not{\hbox{\kern-2pt${#1}$}}}

\newcommand{\drawsquare}[2]{\hbox{%
\rule{#2pt}{#1pt}\hskip-#2pt
\rule{#1pt}{#2pt}\hskip-#1pt
\rule[#1pt]{#1pt}{#2pt}}\rule[#1pt]{#2pt}{#2pt}\hskip-#2pt
\rule{#2pt}{#1pt}}

\newcommand{\Yfund}{\raisebox{-.5pt}{\drawsquare{6.5}{0.4}}}
\newcommand{\Yasymm}{\raisebox
{-3.5pt}{\drawsquare{6.5}{0.4}}\hskip-6.9pt%
                      \raisebox{3pt}{\drawsquare{6.5}{0.4}}%
                     }
\newcommand{\Ysymm}{\Yfund\hskip-0.4pt%
                     \Yfund}

\newcommand\T{\rule{0pt}{2.5ex}}

\newcommand\B{\rule[-1.7ex]{0pt}{0pt}}

\def\drawbox#1#2{\hrule height#2pt
         \hbox{\vrule width#2pt height#1pt \kern#1pt
               \vrule width#2pt}
               \hrule height#2pt}

\def\Asym#1#2{\vcenter{\vbox{\drawbox{#1}{#2}
               \kern-#2pt       
               \drawbox{#1}{#2}}}}

\def\bdot{\huge{\textbf{.}}}

\newcommand {\beq} {\begin{equation}}
\newcommand {\eeq} {\end{equation}}
  \newcommand {\ber}{\begin{eqnarray*}}
  \newcommand {\eer} {\end{eqnarray*}}
\newcommand {\bea}{\begin{eqnarray}}
  \newcommand {\eea} {\end{eqnarray}}

\newcommand{\Dslash}{\,{\raise.15ex\hbox{/}\mkern-12mu D}}

\begin{document}


\begin{titlepage}

\begin{center}
\vspace{1in}
\large{\bf Non-Supersymmetric Brane Configurations,}\\
\large{\bf Seiberg Duality and Dynamical Symmetry Breaking}\\
\vspace{0.4in}
\large{Adi Armoni}\\
\small{\texttt{a.armoni@swan.ac.uk}}\\
\vspace{0.2in}
\emph{Department of Physics, Swansea University}\\ 
\emph{Singleton Park, Swansea, SA2 8PP, UK}\\
\vspace{0.2in}
\emph{Kavli-IPMU, University of Tokyo}\\ 
\emph{Kashiwa, Chiba, Japan}\\
\vspace{0.3in}
\end{center}

\abstract{We consider type IIA brane configurations, similar to those that realize $SO(2N)$ supersymmetric QCD, that include orientifold planes and anti-branes. Such brane configurations lead to $Sp(2N)$ field theories that become supersymmetric in the large-$N$ limit and break supersymmetry upon the inclusion of $1/N$ corrections. We argue that this class of field theories admit Seiberg duality and interpret the potential between branes and orientifolds as field theory phenomena. In particular we find in the magnetic theory a meson potential that leads to dynamical symmetry breaking and a meson condensate similar to the anticipated quark condensate in QCD.}

\end{titlepage}

\section{Introduction}
Understanding the strong coupling regime of QCD remains a notorious challenging problem even after decades of intensive studies. 

In a seminal paper, almost two decades ago, Seiberg argued that the IR of ${\cal N}=1$ super QCD admits two dual descriptions, an $SU(N_c)$ electric theory and an $SU(N_f-N_c)$ magnetic theory. In the so called conformal window, when ${3\over 2} N_c <N_f < 3N_c$, the two theories flow to the same IR fixed point. When $N_c +2 \le N_f \le {3\over 2} N_c$ the electric theory is weakly coupled in the UV and strongly coupled in the IR and the magnetic theory is IR free\cite{Seiberg:1994pq}. The duality statement extends to $SO$ and $Sp$ SQCD.

Seiberg duality provides an insight into the IR degrees of freedom of the strongly coupled theory in terms of weakly coupled fields. One of the surprising outcomes of Seiberg duality is that when  $N_c +2 \le N_f \le {3\over 2} N_c$ the IR of the theory is described not only by massless mesons, but also in terms dual gauge fields and quarks. An interpretation of the dual gauge group as a ``hidden local symmetry'' has been given recently in \cite{Komargodski:2010mc}.

A lot of effort has been made throughout the years to generalize Seiberg duality to a non-supersymmetric theory. One approach is to perturb the electric theory by a relevant operator that breaks supersymmetry and to identify the perturbation in terms of magnetic variables \cite{Aharony:1995zh,Evans:1995rv,Kitano:2011zk}.

 Another approach is to consider ``orbifold'' \cite{Kachru:1998ys} and ``orientifold'' \cite{Sugimoto:1999tx} theories - a class of theories that become planar (large-$N$) equivalent to SQCD \cite{Schmaltz:1998bg,Armoni:1999gc,Armoni:2008gg} in a well defined common sector, and break supersymmetry once $1/N$ corrections are included.

Until recently the main interest in ``orbifold/orientifold theories'' was in the understanding of their large-$N$ equivalence with supersymmetric theories and its implications \cite{Armoni:2003gp}. The finite-$N$ dynamics of these theories remained elusive until recently, where it was argued by Sugimoto \cite{Sugimoto:2012rt}, following ref.\cite{Uranga:1999ib}, that S-duality can be extended to a non-supersymmetric ``orientifold theory'' even at finite-$N$. The breakthrough is due to the understanding of how S-duality acts on a brane configuration that does not preserve supersymmetry. In particular, a repulsive potential between an orientifold plane and branes is interpreted as a Coleman-Weinberg potential that leads to dynamical symmetry breaking of a continuous global symmetry. Additional examples of non-supersymmetric S-dual pairs were given recently in ref.\cite{Hook:2013vza}. Similar ideas and techniques will be used in the present paper. 

In this paper we would like to suggest a Seiberg duality between two ``orientifold field theories''. We will use the string theory embedding and dynamics to support the duality conjecture. Moreover, we will also make use of field theory considerations such as anomaly matching as a supporting evidence for the duality. Note that as in the supersymmetric theory, in the present case the bosonic matter content is uniquely fixed (by either string theory or field theory consideration, as we shall see), hence the global anomaly matching between the proposed dual pair is a stronger evidence with respect to a generic pair of non-supersymmetric theories.

The outcome of the duality is a magnetic theory where the only massless degrees of freedom consist of Nambu-Goldstone mesons. The meson spectrum matches the most naive dynamical symmetry breaking pattern. In particular we will consider a $Sp$ gauge theory with a global $SU(2N_f)$ symmetry. Our analysis supports a breaking of the form 
\beq
 SU(2N_f) \rightarrow Sp(2N_f)
\eeq
and a formation of a meson condensate, similar to the QCD quark condensate. This breaking pattern is anticipated in a QCD like theory due to Vafa-Witten theorem \cite{Vafa:1983tf} and Coleman-Witten analysis  \cite{Coleman:1980mx} at infinite-$N$.

The organization of the paper is as follows: in section 2 we explain the rational behind the duality and write down the matter content of the dual pair. In section 3 we list the global symmetries of the electric and magnetic theories and show in detail how the global anomalies match. In section 4 we describe the string theory origin of the two theories and provide a supporting evidence for the duality. In section 5 we calculate the masses of the squarks in both the electric and magnetic theories. Section 6 is devoted to a calculation of the Coleman-Weinberg potential for the meson field. In section 7 we discuss our results. 

\section{The Electric and Magnetic Field Theories}

We propose a Seiberg duality between a pair of non-supersymmetric gauge
 theories. The field theories that we consider live on non-supersymmetric Hanany-Witten brane configurations \cite{Hanany:1996ie} of type IIA string theory.

From the pure field theoretic point of view we can think about the matter content of our models as a hybrid between $SO(2N_c)$ and $Sp(2N_c)$ SQCD. More precisely, we consider an electric theory with bosons that transform in representations of the $Sp(2N_c)$ SQCD theory and fermions that transform in representations of $SO(2N_c)$ SQCD theory.

Our prime electric theory is given in table \eqref{tableelectric} below. Note in particular that the ``gluino'' transforms in the antisymmetric representation, as if it was the gluino of the $SO(2N_c)$ theory. Note also that in the limit $N_c \rightarrow \infty$, the electric theory become supersymmetric, since in the large-$N_c$ limit there is no distinction between the symmetric and antisymmetric representation. Thus supersymmetry is broken explicitly as a $1/N_c$ effect. We will gain a better understanding of this effect from the string realization of the field theory. 

Both the electric and magnetic theories admit a $SU(2N_f)\times U(1)_R$ global symmetry. Note that $U(1)_R$ is simply a name for the axial symmetry, borrowed from the supersymmetric model.

\begin{table}[!ht]
\begin{center}
\begin{tabular}{|c|c|cc|}
\hline
\multicolumn{4}{|c|} {Electric Theory} \\ 
\hline \hline
 & $Sp(2N_c)$ & $SU(2N_f)$  & $U(1)_R$ \\
\hline
 $A_\mu$ & \Ysymm & \bdot & 0 \\
& $N_c(2N_c+1)\B$ & & \\
\hline
$\lambda$ & $\T\Yasymm\B$   &  \bdot  &  1  \\
& $N_c(2N_c-1)\B$ & & \\
                   \hline
$\Phi$   & $\Yfund$ & $\Yfund$ & $\T\frac{N_f-N_c+1}{N_f}$ \\
&   $2N_c$  &   $2N_f$ & \\
\hline
$\Psi$ & $\Yfund$ & $\Yfund$ & $\T\frac{-N_c+1}{N_f}$ \\
&   $2N_c$  &   $2N_f$  & \\                   
\hline
\end{tabular}
\caption{\it The matter content of the electric theory.}
\label{tableelectric}
\end{center}
\end{table}

Let us consider the magnetic theory. Its matter content is given in table \eqref{tablemagnetic} below. It is obtained by changing the representation of the gluino in the $Sp$ magnetic supersymmetric theory from symmetric to antisymmetric and by replacing the representation of the mesino from antisymmetric to symmetric. 

Note that $\tilde N_c = N_f-N_c +2$, as in the duality between a supersymmetric $SO$ pair.

\begin{table}[!ht]
\begin{center}
\begin{tabular}{|c|c|cc|}
\hline
\multicolumn{4}{|c|} {Magnetic Theory} \\ 
\hline \hline
 & $\T Sp(2\tilde N_c)$ & $SU(2N_f)$  & $U(1)_R$ \\
\hline
 $a_\mu$ & \Ysymm & \bdot & 0 \\
& $\tilde N_c(2\tilde N_c+1)\B$ & & \\
\hline
$l$ & $\T\Yasymm\B$   &  \bdot  &  1  \\
& $\tilde N_c(2\tilde N_c-1)\B$ & & \\
                   \hline
$\phi$   & $\Yfund$ & $\bar{\Yfund}$ & $\T\frac{N_c-1}{N_f}$ \\
&   $2\tilde N_c$  &   $2\bar{N_f}$ & \\
\hline
$\psi$ & $\Yfund$ & $\bar{\Yfund}$ & $\T\frac{N_c-N_f-1}{N_f}$ \\
&   $2\tilde N_c$  &   $2\bar{N_f}$  & \\  
         \hline
 $M$ & \bdot & \Yasymm & $\T\frac{2N_f-2 N_c +2}{N_f}$  \\
& & $N_f(2N_f-1)$  & \\
\hline
$\chi$ & \bdot & $\T\Ysymm\B$   &  $\T\frac{N_f-2N_c +2}{N_f}$  \\
& & $N_f(2N_f+1)$ & \\          
\hline
\end{tabular}
\caption{\it The matter content of the magnetic theory. $\tilde N_c=N_f-N_c+2$.}
\label{tablemagnetic}
\end{center}
\end{table}

We will argue that the electric and the magnetic form a dual pair. Note in particular that in the Veneziano limit, $N_c \rightarrow \infty$, with $N_f/N_c$ fixed, this is a simple statement, since in the large $N$ limit the theories become supersymmetric. Our statement is about the finite $N$ theory. A weak version of the statement, that we will adopt throughout the paper, is that we include only the leading $1/N$ correction, such that supersymmetry breaking is a small perturbation.

An important remark is about the couplings in the electric and magnetic theories. When the theory is supersymmetric there are relations between the various couplings that appear in the Lagrangian. In the absence of supersymmetry one has to list the relations between the various couplings. We will simply use the same relations between couplings as in the supersymmetric case. We expect that when $N$ is large the supersymmetric ratios between the couplings are modified by a small $1/N$ correction that will not affect the IR theory.

\section{Anomaly matching}

A consistency check of our proposal, that we can always perform irrespectively of supersymmetry, is 't Hooft anomaly matching.

We will match the global anomalies for $SU(2N_f)^3$, $SU(2N_f)^2 U(1)_R$, $U(1)_R$ and $U(1)^3_R$ in table \ref{tab:anomaly} below. We use the notation $\tilde{N_c}=N_f-N_c+2$
and the terms in each box are ordered as (gluino) + (quarks) in the electric theory 
and (gluino) + (quarks) + (mesino) in the magnetic theory. 
$d^2(R)\delta^{ab}$ and $d^3(R)d^{abc}$ for the representation $R$ are respectively 
the traces ${\rm tr}_R \, T^a T^b$, ${\rm tr}_R \, T^a \{ T^b, T^c\}$.
In table \ref{tab:anomaly} we make use of the following relations:
\beq
\label{eviaa}
d^2(\Ysymm)=(2N_f+2)d^2(\Yfund)~, ~ ~ 
d^3(\Ysymm)=(2N_f+4)d^3(\Yfund)
~.
\eeq

Note that the matching works as the matching of anomalies in the supersymmetric $SO(2N_c)$ case. This is not surprising, since 
the fermions in our model carry the same representations as the fermions 
in $SO(2N_c)$ SQCD.

\begin{table}[!ht]
\begin{center}
\begin{tabular}{|l|l|l|}
\hline
 & Electric & Magnetic \\
\hline \hline
& & \\
$SU(2N_f)^3$ & 
$0 + 2N_cd^3(\Yfund)$ & $0 + \tilde{2N_c}(-d^3(\Yfund))+d^3(\Ysymm)$\\
& $=2N_cd^3(\Yfund)$ & $=2N_cd^3(\Yfund)$\\
& & \\
\hline \hline
& & \\
$SU(2N_f)^2U(1)_R$ &
$0 + 2N_c\left(\frac{-N_c+1}{N_f}\right)d^2(\Yfund)$ & $0 
+ \tilde{2N_c}\left(\frac{N_c-N_f-1}{N_f}\right)d^2(\Yfund)+$\\
& $=\frac{-2N_c^2+2N_c}{N_f}d^2(\Yfund)$ 
& $+\left(\frac{N_f-2N_c+2}{N_f}\right)d^2(\Ysymm)=\frac{-2N_c^2+2N_c}{N_f}d^2(\Yfund)$\\
& & \\
\hline \hline
& & \\
$U(1)_R$ &
$(2N_c^2-N_c)+$ 
& $(\tilde{2N_c^2}-\tilde{N_c})+4\left(\tilde{N_c}N_f\frac{N_c-N_f-1}{N_f}\right)$+\\
& $+4\left(N_cN_f\frac{-Nc+1}{N_f}\right)$
& $(2N_f^2+N_f)\left(\frac{N_f-2N_c+2}{N_f}\right)$\\
& $=-2N_c^2+3N_c$ & $=-2N_c^2+3N_c$\\
& & \\
\hline \hline
& & \\
$U(1)_R^3$ &
$(2N_c^2-N_c)+$ 
& $(\tilde{2N_c^2}-\tilde{N_c})+4\left[\tilde{N_c}N_f\left(\frac{N_c-N_f-1}{N_f}\right)^3\right]+$ \\
& $+4\left[N_cN_f\left(\frac{-Nc+1}{N_f}\right)^3\right]$
& $\left[(2N_f^2+N_f)\left(\frac{N_f-2N_c+2}{N_f}\right)^3\right]$\\
& $=N_c\left(2N_c-1-4\frac{(N_c-1)^3}{N_f^2}\right)$ 
& $=N_c\left(2N_c-1-4\frac{(N_c-1)^3}{N_f^2}\right)$\\
& & \\
\hline
\end{tabular}
\caption{\it 't Hooft anomaly matching.}
\label{tab:anomaly}
\end{center}
\end{table}

The matching of global anomalies is very encouraging. Of course since anomalies concern only the fermionic sector of the theory one may wonder whether the matching fixes the bosonic matter content. In the supersymmetric case we know that it is enough to fix either the bosonic or the fermionic content of the theory. This is not the case in a generic non-supersymmetric theory, but it is the case for the present electric and magnetic theories. The entire matter content of the above theories is fixed by certain brane configurations. Brane dynamics also fixes the rank of the dual gauge group. From the field theoretic point of view we can claim that the matter content is determined by the principle that the theory is a hybrid of bosons that transform in $Sp$ SQCD and fermions that transform in $SO$ SQCD.

\section{Brane configurations that include O4 planes and anti D branes}

In order to obtain an intuition about the class of non-supersymmetric field theories and the proposed Seiberg duality between the electric and magnetic theories, let us consider their string theory origin.

The class of theories that we consider are called ``orientifold field theories''. These theories live on a brane configuration that consists of an orientifold plane and {\it anti-branes} \cite{Sugimoto:1999tx}. These brane configurations break supersymmetry, but the supersymmetry breaking effect is suppressed by $1/N$. The reason is that the M\"{o}bius amplitude, that leads to supersymmetry breaking, contributes to the free energy as ${\cal O}(N)$ while the leading annulus diagram contribution is ${\cal O}(N^2)$. The fact that supersymmetry breaking is a $1/N$ effect is a good starting point. It essentially means that in the large-$N$ limit we consider a small perturbation around the supersymmetric theory, where holomorphicity leads to solid non-perturbative results.

Let us focus on the brane configuration that gives rise to the electric theory. It is identical to the brane configuration that realizes $SO(2N)$ SQCD, except 
that the D4-branes are replaced by {\it anti} D4-branes. The brane configuration is depicted in figure \eqref{electric-figure} below.

\begin{figure}[!ht]
\centerline{\includegraphics[width=8cm]{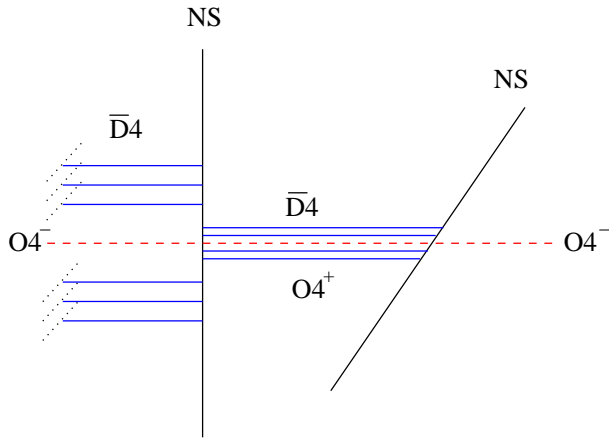}}
\caption{\footnotesize The type IIA brane configuration that realizes the electric theory.} \label{electric-figure}
\end{figure}

The brane configuration consists of $N_c$ anti D4-branes and their mirror branes. The ``color-color'' strings, in the presence of the $O4^+$ plane, give rise to a gluon in the adjoint (two-index symmetric) and a gluino that transforms in the two-index antisymmetric representation of the $Sp(2N_c)$ group \cite{Sugimoto:1999tx}. In addition there are ``color-flavor'' strings that lead to $N_f$ quarks and squarks. An important comment is that due to the presence of the orientifold $O4^-$ plane the brane configuration realizes an $SO(2N_f)$ subgroup of the full $SU(2N_f)$ global symmetry of the theory in table \eqref{tableelectric}. We will discuss this matter in more detail shortly, when we will describe the magnetic theory. The matter content of the electric theory that lives on the brane configuration is listed in table \eqref{tableelectric2} below.

\begin{table}[!ht]
\begin{center}
\begin{tabular}{|c|c|cc|}
\hline
\multicolumn{4}{|c|} {Electric Theory on the Brane} \\ 
\hline \hline
 & $Sp(2N_c)$ & $SO(2N_f)$  & $U(1)_R$ \\
\hline
 $A_\mu$ & \Ysymm & \bdot & 0 \\
& $N_c(2N_c+1)\B$ & & \\
\hline
$\lambda$ & $\T\Yasymm\B$   &  \bdot  &  1  \\
& $N_c(2N_c-1)\B$ & & \\
                   \hline
$\Phi$   & $\Yfund$ & $\Yfund$ & $\T\frac{N_f-N_c+1}{N_f}$ \\
&   $2N_c$  &   $2N_f$ & \\
\hline
$\Psi$ & $\Yfund$ & $\Yfund$ & $\T\frac{-N_c+1}{N_f}$ \\
&   $2N_c$  &   $2N_f$  & \\                   
\hline
\end{tabular}
\caption{\it The matter content of the electric theory that lives on the brane configuration.}
\label{tableelectric2}
\end{center}
\end{table}

In order to obtain the magnetic theory we proceed as in \cite{Elitzur:1997fh} and \cite{Evans:1997hk}. We swap the NS5 branes. In the presence of an orientifold plane two anti D4-branes and their mirrors are created as color branes. It therefore leads to a theory based on a $Sp(2\tilde N_c)$ gauge group, with $\tilde N_c = N_f - N_c +2$. The theory is depicted in figure \eqref{magnetic-figure} below.

\begin{figure}[!ht]
\centerline{\includegraphics[width=8cm]{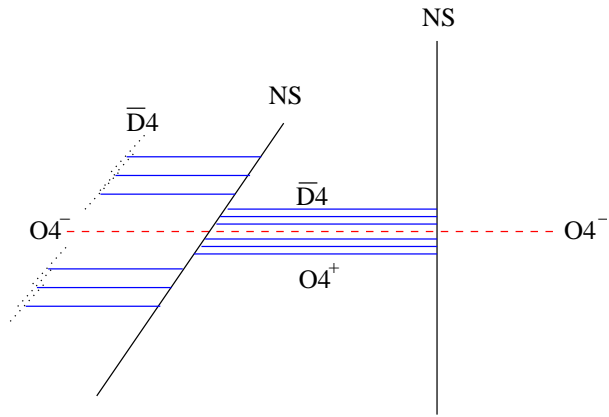}}
\caption{\footnotesize The type IIA brane configuration that realizes the electric theory.} \label{magnetic-figure}
\end{figure}

We obtain a magnetic theory with a gluon that transforms in the adjoint representation and a ``gluino'' that transforms in the two-index antisymmetric representation of the group $Sp(2\tilde N_c)$, due to ``color-color'' strings. In addition we have $N_f$ quarks and squarks, due to ``color-flavor'' strings. Finally we have a meson and a mesino, due to ``flavor-flavor'' strings. The meson transforms in the two-index antisymmetric representation and the mesino transforms in the two-index symmetric representation of $SO(2N_f)$ group. The reason that the global symmetry is $SO(2N_f)$ is that the strings cross the orientifold $O4^-$ plane.  The matter content of the magnetic theory that lives on the brane configuration is listed in table \eqref{tablemagnetic2} below.

\begin{table}[!ht]
\begin{center}
\begin{tabular}{|c|c|cc|}
\hline
\multicolumn{4}{|c|} {Magnetic Theory on the Brane} \\ 
\hline \hline
 & $\T Sp(2\tilde N_c)$ & $SO(2N_f)$  & $U(1)_R$ \\
\hline
 $a_\mu$ & \Ysymm & \bdot & 0 \\
& $\tilde N_c(2\tilde N_c+1)\B$ & & \\
\hline
$l$ & $\T\Yasymm\B$   &  \bdot  &  1  \\
& $\tilde N_c(2\tilde N_c-1)\B$ & & \\
                   \hline
$\phi$   & $\Yfund$ & $\Yfund$ & $\T\frac{N_c-1}{N_f}$ \\
&   $2\tilde N_c$  &   $2N_f$ & \\
\hline
$\psi$ & $\Yfund$ & $\Yfund$ & $\T\frac{N_c-N_f-1}{N_f}$ \\
&   $2\tilde N_c$  &   $2N_f$  & \\  
         \hline
 $M$ & \bdot & \Yasymm & $\T\frac{2N_f-2 N_c +2}{N_f}$  \\
& & $N_f(2N_f-1)$  & \\
\hline
$\chi$ & \bdot & $\T\Ysymm\B$   &  $\T\frac{N_f-2N_c +2}{N_f}$  \\
& & $N_f(2N_f+1)$ & \\          
\hline
\end{tabular}
\caption{\it The matter content of the magnetic theory that lives on the brane configuration.}
\label{tablemagnetic2}
\end{center}
\end{table}

Note that the interpolation between the electric and magnetic theories does not rely on supersymmetry. Each step is on equal footing with the corresponding step in the SQCD case. The main question, which is crucial, is why the interpolation should lead to a Seiberg dual. The same question could, in fact, be raised even in the supersymmetric case. A partial answer, restricted to holomorphic data, is given in \cite{Hori:1998iw}, where Seiberg duality is understood as two weakly coupled limits of a single configuration in M-theory. In the present case we do not have a convincing answer to this question and for this reason we cannot claim that we have a proof of Seiberg duality. We can only propose this duality and test it. We wish, however, to note other cases of Seiberg dual pairs with two supercharges or no supersymmetry at all \cite{Armoni:2008gg,Armoni:2009vv}. We learn that the ``swapping branes'' argument leads to Seiberg duality even for theories with less than four supercharges.

Due to the lack of supersymmetry there will be forces between the orientifold plane and the anti-branes. In the next section we will analyze those interactions and will give them a field theory interpretation. It turns out that effects in the string theory side capture important physics in gauge dynamics and vice versa.

\section{One loop effects in the electric and magnetic theories and their string theory interpretation}

Since we are interested in the field theories of tables \eqref{tableelectric2} and \eqref{tablemagnetic2} that live on the brane configurations in figures \eqref{electric-figure} and \eqref{magnetic-figure}, we will focus our attention on those field theories. Our analysis, however, also applies to the original theories in tables \eqref{tableelectric} and \eqref{tablemagnetic}.

In the limit $N_c \rightarrow \infty$ the theory acquires supersymmetry and it admits a moduli-space of vacua and massless scalars. When $1/N_c$ corrections are included, scalars acquire either a positive mass$^2$ or a negative mass$^2$ (a tachyon). The potential for the various scalars will be the most important ingredient in the analysis. It will be given an interpretation of a potential between the orientifold plane and branes.

\subsection{Squark potential in the electric theory}

The squark in the electric theory couples to the gluon and to the gluino. Both run in the loop and both lead to quadratic divergences. In the supersymmetric case there is a perfect cancellation between the contribution of the gluon and the gluino, hence the scalar remains massless.  This is not the case at finite $N_c$.

Let us consider the one-loop contribution to the squark mass, as depicted in figure \eqref{mass} below.

\begin{figure}[!ht]
\centerline{\includegraphics[width=9cm]{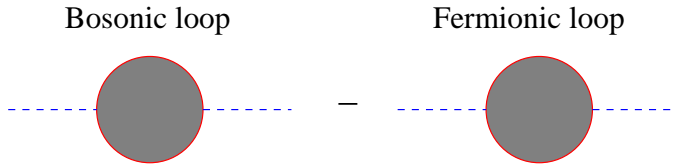}}
\caption{\footnotesize Perturbative contributions to the squark mass. The Bosonic loop is proportional to $2N+1$, while the fermionic loop is proportional to $2N-1$.} 
\label{mass}
\end{figure}

 The contribution from a bosonic one-loop is as in the 
supersymmetric theory
\beq
 +g_e^2 (2N_c+1) \int {d^4 p\over (2\pi)^4} {1\over p^2} = +c g_e^2 (2N_c+1) \Lambda ^2 \, ,
\eeq
with $g_e$ the electric gauge coupling and $\Lambda$ the UV cut-off.

The fermionic one-loop contribution (with a quark and a gluino running in the loop) is
\beq
 -g_e^2 (2N_c-1) \int {d^4 p\over (2\pi)^4} {1\over p^2} = -c g_e^2 (2N_c-1) \Lambda ^2 \, .
\eeq
The generated mass for the squark is therefore
\beq
M^2_{\Phi} = cg_e^2 \left \{ (2N_c+1) - (2N_c-1) \right \} \Lambda ^2 = 2c{g_e^2 N_c \over N_c} \Lambda ^2 \, ,
\eeq 
where $\Lambda$ is interpreted as the UV cut-off of the theory. In field theory quadratic divergences can be removed order by order in perturbation theory as part of the renormalization procedure. Due to the embedding in string theory with $\Lambda ^2 \sim {1\over \alpha '}$ as the natural UV cut-off, we wish to give the generated mass a physical interpretation. We argue that the scalars acquire a mass and decouple from the low-energy dynamics. Below a certain energy scale the physics will be described by an $Sp(2N_c)$ gauge theory coupled to a single fermion in the antisymmetric representation and $N_f$ fundamental quarks.

In an $Sp(2N_c)$ theory with $N_f$ quarks the global $SO(2N_f)$ is expected to break, due to a formation of a quark condensate $\langle \Psi_{[i} \Psi_{j]} \rangle$. The most naive scenario is
\beq
SO(2N_f) \rightarrow U(N_f) \,.
\label{dsb}
\eeq
As we shall see, the magnetic theory supports such a scenario. Note that the above dynamical breaking \eqref{dsb} {\it must} occur in the limit of large $N_c$ with fixed $N_f$ due to Coleman and Witten \cite{Coleman:1980mx}. Moreover, in a theory where the scalars are heavier than the QCD scale, the breaking \eqref{dsb} is very likely to occur, due to Vafa-Witten theorem \cite{Vafa:1983tf} that forbids a breaking of a vector symmetry.

It is interesting to ask what would happen if $M^2 _{\Phi} <0$. This is the case when we place $O4^-$ plane between the NS5 branes and the theory is an $SO(2N_c)$ gauge theory coupled to $N_f$ quarks with $Sp(2N_f)$ global symmetry. When the squark mass$^2$ is negative it acquires a vev $v^a_i = v^a \delta ^a _i$, ``color-flavor locking'' occurs and both gauge and flavor symmetry are broken. Such an effect will be captured in the brane system by a reconnection of color and flavor branes and their repulsion from the orientifold plane.

\subsection{Squark potential in the magnetic theory}

The calculation of the squark mass in the magnetic theory
 is similar to the corresponding calculation in the electric theory, but it is somewhat more subtle. The reason is that the squark is coupled both to the gluon and gluino via a gauge interaction (with $g_m$ the magnetic gauge coupling) and to the meson and mesino via a Yukawa interaction (with $y$ the Yukawa coupling). We thus have a bosonic loop proportional to $g_m^2(2\tilde N_c+1)$, a fermionic loop proportional to $g_m^2(2\tilde N_c -1)$, a bosonic loop contribution proportional to $y^2 (2N_f-1)$ and a fermionic loop proportional to $y^2(2N_f+1)$.
Altogether the various contributions to the magnetic squark mass are
\beq
cg_m^2 \{ (2\tilde N_c+1)-(2\tilde N_c -1) \} \Lambda ^2 +cy^2 \{ (2N_f-1)-(2N_f+1) \} \Lambda ^2 \, .
\eeq
Thus, similarly to the calculation in the previous section
\beq
M^2_{\phi} =2c (g_m^2-y^2) \Lambda ^2 \, .
\eeq
It is therefore crucial to know which one of the couplings is larger, $g_m^2$ or $y^2$.

Without the knowledge of the relation between $g_m^2$ and $y^2$ Seiberg duality is not complete. It was shown in \cite{Oehme:1998yw} (see also \cite{Gardi:1998ch}) that if $g_m^2/y^2$ admits a certain ratio, the two couplings share the same beta function up to two-loop order. The ratio reduces in the Veneziano large-$N_c$ limit to
\beq
{g_m^2 \over y^2} = 3 {N_f\over N_c}-1  \, ,
\eeq
and in particular when $N_c < N_f < {3\over 2} N_c$, $g_m^2 > y^2$, therefore the magnetic squark becomes massive (as the squark in the electric theory).

The generated mass of the squark is $M^2_{\phi} \sim +g_m^2 \Lambda ^2$. We will discuss the implication of this fact in the following section.

\section{The meson potential and its string theory interpretation}

In this section we discuss the Coleman-Weinberg potential for the meson
 field and its implication on the dynamics of the electric field theory. The magnetic theory is rather involved and therefore it is difficult to carry out a reliable calculation. For this reason we will limit ourselves to a one-loop calculation which can be trusted only for small values of the meson's vev. The reason is that in an IR free theory, the coupling becomes stronger as the energy scale becomes higher. When a small vev is introduced the coupling freezes at long distances and stops running at weak coupling. 

In addition to the generated one-loop meson potential, the (large-$N$) theory inherits a potential from the supersymmetric theory, due to the generated superpotential \cite{Intriligator:2006dd} 
\beq
 W=(N_c-N_f)\left ({\det \, M \over \Lambda_{QCD} ^{3N_c -N_f}} \right)^{1\over N_f-N_c} \, .
\eeq
 Upon the inclusion of $1/N$ corrections, when supersymmetry is broken, the effect of this non-perturbative superpotential is not important for ${\langle M_{[ij]} \rangle \over \Lambda_{QCD}^2} \ll 1$. It will not alter our conclusion that the global $SO(2N_f)$ symmetry breaks dynamically.

In addition we will also discuss the interpretation of the meson potential as a potential between the branes and the orientifold of the magnetic configuration. As we shall see, dynamical symmetry breaking can be understood due to a repulsion between the branes and the orientifold plane.

\subsection{The Coleman-Weinberg potential} 

In the previous section we learned that the magnetic squark acquires a mass $\sim g_m^2 \Lambda ^2$. This is a small mass in the large-$N_f$ limit, since
$M^2 \sim {g_m^2 N_f \over N_f} \Lambda ^2$. Note that the magnetic quark remains massless. The meson field will acquire a non-trivial potential due to the Yukawa interaction with the massive squark and the massless quark. The Coleman-Weinberg potential for the vev $\langle M_{[ij]}\rangle \equiv m_{[ij]}$ takes the form
\bea 
& & 
 V(\{ m_{[ij]} \})= \\
& &
 \tilde N_c \left ( {\rm tr} \int {d^4 p \over (2\pi)^4} \log (p^2 +y^2 (mm^\dagger) + g_m^2 \Lambda ^2) - {\rm tr}\int {d^4 p \over (2\pi)^4} \log (p^2 +y^2 (mm^\dagger)) \right )\, . \nonumber
\eea
We will use \cite{Tseytlin:1999tp}
\bea
& & F(\mu^2)=\int {d^4 p \over (2\pi)^4} \log (p^2 +\mu^2) = \\
& & -{1\over (4\pi)^2} \int _{1\over \Lambda ^2} ^\infty {dt \over t^3} \exp (-t \mu ^2) = c_0 \Lambda ^4 + c_1 \Lambda ^2 \mu ^2 + {1\over 4\pi^2} \mu ^4 \log {\mu ^2 \over \Lambda ^2} \nonumber
\eea
and
\beq
 V(\{ m_{[ij]} \})=  \tilde N_c{\rm tr}\left ( F(y^2 mm^\dagger +g_m^2 \Lambda ^2)-F(y^2 mm^\dagger) \right ) \, ,
\eeq
to arrive at
\beq
\hat V(\{ \hat m_{[ij]}\}) = {1 \over 4\pi ^2} {\rm tr} \left \{ (\hat m {\hat m}^\dagger +g_m ^2)^2 \log (\hat m {\hat m}^\dagger + g_m^2)- (\hat m {\hat m}^\dagger) ^2 \log (\hat m {\hat m}^\dagger) \right \} 
\label{potential}
\eeq
where $\hat V = V/ (\tilde N_c \Lambda ^4)$ and $\hat m =y m / \Lambda$. 

\begin{figure}[!ht]
\centerline{\includegraphics[width=8cm]{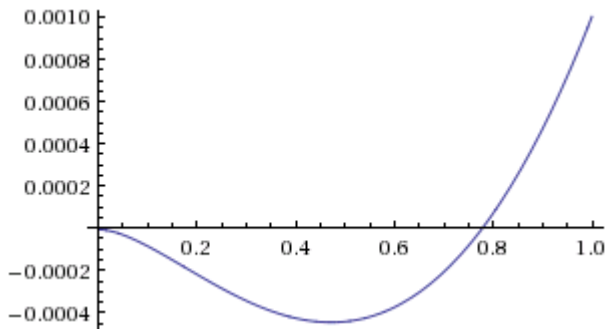}}
\caption{\footnotesize The potential $4\pi ^2 \hat V$ for the meson field. In this figure we used $g_m^2=0.001$. The minimum is at $\hat m \hat m^\dagger =\exp (-3/2)$.} 
\label{graph}
\end{figure}

When $g_m^2 \ll 1$ (this is indeed the case, since $N_f$ is large and $g_m^2  N_f$ is kept fixed), the function $\hat V$ admits a unique minimum at $\hat m \hat m^\dagger=\exp (-3/2)$, which is independent of $g_m^2$. The function $\hat V$ is plotted in figure \eqref{graph}.

And thus the vacuum solution for the meson matrix takes the form
\beq
\langle M_{[ij]} \rangle =
\begin{pmatrix}
 0 & m &  &  &  & & \\
-m & 0 &  &  &  & & \\
   & & 0 & m &  & & \\
   & & -m & 0 & & & \\
   & &    &   & \ddots & & \\
   & &  & & & 0 & m \\
   & &  & & & -m & 0 
\end{pmatrix}
\eeq
with $ym = \exp (-3/4) \Lambda$ and $\Lambda$ is the cut-off of the magnetic theory.

The one-loop analysis of the Coleman-Weinberg potential in the magnetic theory yields a vacuum solution where the $SO(2N_f)$ is dynamically broken to $U(N_f)$. As a result there are ${1\over 2} 2N_f (2N_f-1) - N_f^2 = N_f^2 - N_f$ massless Nambu-Goldstone bosons in the coset $SO(2N_f)/U(N_f)$. They correspond to flat directions of the potential. The rest of the mesons, that correspond to non-flat directions, acquire a mass $M^2 \sim {\Lambda ^2 \over N_c}$. In additional to the breaking of the global $SO(2N_f)$ symmetry, the $U(1)_R$ symmetry also gets broken. 

Let us discuss the corresponding condensate in the electric theory. If we use the dictionary of the supersymmetric theory 
\beq
M_{[ij]} =  \Phi ^a _{[i} \Phi^a _{j]} \, ,
\eeq 
where $\Phi$ is the electric squark. The equations of motion for the massive field $\Phi ^a_i$ relate it to $\lambda \Psi$ as follow
\beq
\Phi ^a _i \sim \lambda ^a _b \Psi ^b _i \, ,
\eeq
hence the meson condensate can be identified with the four fermion condensate
\beq
\langle M_{[ij]} \rangle \sim \langle \lambda \Psi _{[i} \lambda \Psi _{j]} \rangle
\label{condensate}
\eeq
A consistency check of the above identification \eqref{condensate} is that 
both the meson operator and the four fermion electric operator have the same $U(1)_R$ charge, with $R={2(N_f - N_c+1) \over N_f}$. Dynamical symmetry breaking is thus understood as due to quark condensation \eqref{condensate}, similarly to the chiral quark condensate formation in QCD.

Note that while the generated potential \eqref{potential} is a $1/N$ effect, the meson condensate is {\it not} a $1/N$ effect. This is what we expect from the electric theory: the breaking of supersymmetry selects a vacuum where the global flavor symmetry is broken.

Finally, let us comment on the fate of the $Sp(2\tilde N_c)$ gauge theory. As the meson condenses, both the squark and the quark acquire a mass due to the superpotential $W={1\over \mu}Mqq$. The color and flavor theories will decouple. The $Sp(2\tilde N_c)$ theory is expected to confine and to exhibit a mass gap, similar to pure ${\cal N}=1$ Super Yang-Mills theory. The glueballs of the color theory are massive and hence decouple from the IR theory. Therefore the only massless fields of the magnetic theory are the $N_f^2-N_f$ Nambu-Goldstone bosons associated with the breaking of the $SO(2N_f)$ flavor symmetry and an additional Nambu-Goldstone boson associated with the breaking of the $U(1)_R$ symmetry. 

\subsection{Brane picture interpretation}

Let us provide an interpretation of the potential \eqref{potential} in terms of brane dynamics. We focus on the magnetic brane configuration \eqref{magnetic-figure}.

\begin{figure}[!ht]
\centerline{\includegraphics[width=5cm]{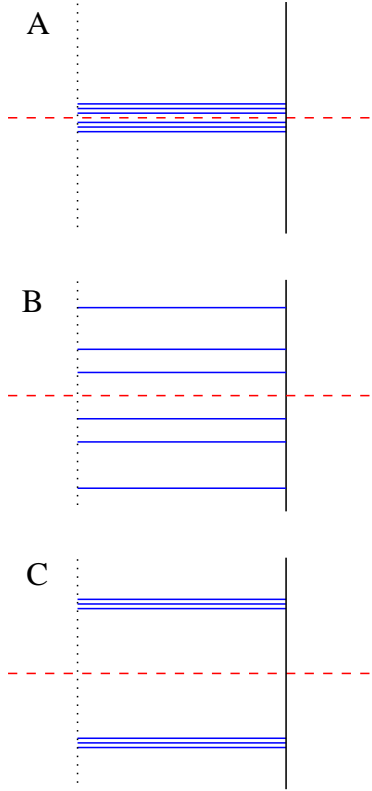}}
\caption{\footnotesize Three possible vacuum configurations. The D4 branes are represented by solid blue lines which end on an NS5 brane (black line) and D6 branes (dotted lines). The orientifold plane is represented by a red dashed line. At the top (A) configuration the branes coincide and sit on top of the orientifold plane. In the middle (B) configurations the branes split away from the orientifold. In the bottom configuration (C), which is selected by the one-loop Coleman-Weinberg potential \eqref{potential}, the branes coincide and sit away from the orientifold plane.}
 \label{vacuum}
\end{figure}

The vev's of the meson field can be interpreted as distances between the orientifold plane and the D4 branes. In particular when $\langle M_{[ij]} \rangle =0$ the D4 branes (and their mirrors) coincide and ``sit'' on top of the orientifold planes. This is depicted in fig.\ref{vacuum}a (top) below.

 The field theory interpretation is that at this point the vacuum admits an $SO(2N_f)$ symmetry. Another possibility, depicted in fig.\ref{vacuum}b (middle) is when
\beq
\langle M_{[ij]} \rangle =
\begin{pmatrix}
 0 & m_1 &  &  &  & & \\
-m_1 & 0 &  &  &  & & \\
   & & 0 & m_2 &  & & \\
   & & -m_2 & 0 & & & \\
   & &    &   & \ddots & & \\
   & &  & & & 0 & m_{N_f} \\
   & &  & & & -m_{N_f} & 0 
\end{pmatrix}
\eeq
namely the D4 branes (and their mirrors) separate and ``sit'' at distinct points away from the orientifold plane. In this case the interpretation is that the vacuum admits $U(1)^{N_f}$ flavor symmetry. 

The potential \eqref{potential} selects a solution where all the D4 branes (and their mirror) coincide and ``sit'' away from the orientifold plane. This is depicted in fig.\ref{vacuum}c (bottom). This vacuum configuration corresponds to a $U(N_f)$ symmetry.

We may interpret the potential \eqref{potential} as the potential between the 
 flavor branes and the orientifold plane. Effectively, the branes are repelled away from the orientifold and find a minimum at a certain position $\langle m \rangle$. This is very similar to the scenario of ref.\cite{Sugimoto:2012rt}, where the anti-D3 branes of the magnetic theory had been repelled from the orientifold O3 place, resulting in dynamical symmetry breaking of the form $SU(4) \rightarrow SO(4)$.

We can also interpret the open strings between anti D4 branes as massless and massive mesons. At the origin, in fig.\eqref{vacuum}a, there are $2N_f(2N_f -1)$ possible open strings that correspond to the various entries of the complex meson matrix $M_{[ij]}$. These strings split into two kinds at the vacuum configuration of \eqref{vacuum}c: ``short strings'' and ``long strings''. There are $2N_f^2$ ``short strings'' that connect branes on one side of the orientifold. $2N_f^2-1$ of these strings are massive and one is massless. The ``center of mass'' $U(1)$ corresponds to the Nambu-Goldstone boson associated with the spontaneously broken $U(1)_R$ symmetry. In addition there are  $2(N_f^2-N_f)$ long strings that cross the orientifold plane. Half of the long strings correspond to Nambu-Goldstone bosons. If the theory on the flavor branes was a gauge theory, $N_f^2-N_f$ of the long strings would correspond to W-bosons whose masses are $M_W=gv$ (with $g$ being the gauge coupling of the would-be $SO(2N_f)$ gauge theory). That could have been achieved by replacing the D6 branes with an NS5 brane. However, we are interested in a theory where the $SO(2N_f)$ symmetry is not gauged and $g \rightarrow 0$. In this limit the W-bosons become massless Nambu-Goldstone bosons. 

\section{Discussion}

In this paper we proposed a duality between a pair of ``orientifold field theories''. The main support for our proposal is the embedding in string theory and the matching of global anomalies. In the large-$N$ limit the theories become supersymmetric and hence our proposal in this limit becomes the standard Seiberg duality between electric and magnetic $Sp$ SQCD.

The theories that we considered admit either $SU(2N_f)$ global symmetry, or a reduced $SO(2N_f)$ symmetry when the theories are realized on a brane configuration. The one-loop potential that results from the duality leads to a breaking $SO(2N_f) \rightarrow U(N_f)$ for the theory on the brane. For the theory \eqref{tablemagnetic} the same potential \eqref{potential} breaks $SU(2N_f) \rightarrow Sp(2N_f)$.   

We would also like to emphasize that we cannot prove our proposal, but the emerging picture is encouraging. If we consider the electric theory in \eqref{tableelectric} at finite $N_c$, we anticipate that the global $SU(2N_f)$ symmetry breaks dynamically to $Sp(2N_f)$. This scenario is compatible with both the Coleman-Witten argument and with Vafa-Witten theorem \cite{Vafa:1983tf}. This is indeed the result of the one-loop analysis \eqref{potential}. In addition, the GMOR relation $f^2_{\pi}M^2_{\pi}=m_q \langle \bar \psi \psi \rangle$ is expected to emerge naturally from non-supersymmetric Seiberg duality, due to the superpotential $W=m_q M +{1\over \mu} M qq$ that gives the pion a mass, $M^2 \sim m_q$. The GMOR relation and other phenomenological implications, such as the $\eta'$ mass, deserves further investigation.

The outcome of this paper and \cite{Sugimoto:2012rt} as well as previous works, is that the breaking of supersymmetry in ``orientifold field theory'' is a mild $1/N$ effect: the large-$N$ theory inherits supersymmetric properties, such as S-duality or Seiberg duality.

A possible future direction is to consider other brane configurations that admit ${\cal N}=1$ supersymmetry and Seiberg duality and to replace branes by anti-branes. An interesting class of such theories was introduced recently in refs.\cite{GarciaEtxebarria:2012qx,Garcia-Etxebarria:2013tba}.

Another future application of this program is to consider ``orientifold field theories'' analogous to ${\cal N}=2$ super Yang-Mills. Such theories live on a Hanany-Witten brane configuration that consists of an orientifold plane, parallel NS5 branes and anti D4 branes. It is interesting to understand what happens to the Seiberg-Witten curve and to the IR theory upon the inclusion of $1/N$ corrections. 

Finally, we would like to mention that we carried out a similar analysis for a $U(N_c)$ QCD-like theory that lives on a type 0' brane configurations \cite{Armoni:2013kpa}. In that case, Seiberg duality suggests chiral symmetry breaking pattern of the form $SU_L(N_f) \times SU_R(N_f) \rightarrow SU_V(N_f)$ with the corresponding $N_f^2-1$ pions, as in real QCD!

\vskip 0.5cm
{\it \bf Acknowledgments.} I wish to thank Kavli-IPMU for a warm hospitality while this work has been carried out. I am indebted to Shigeki Sugimoto for collaboration and for numerous insightful discussions.


\end{document}